\documentclass[iop]{emulateapj}                                            %

\usepackage{epsfig}

\newcommand{\be}{\begin{equation}}
\newcommand{\ee}{\end{equation}}
\newcommand{\ba}{\begin{eqnarray}}
\newcommand{\ea}{\end{eqnarray}}

\tabletypesize{\small}

\slugcomment{Appeared in ApJ 742, 17, (2011)}
\shorttitle{Mas and Radius of XTE J1807}

\begin{document}
\title{Constraints on the Mass and Radius of the Neutron Star XTE J1807-294}
\author {Denis A. Leahy\altaffilmark{1}, Sharon M. Morsink\altaffilmark{2}
\& Yi Chou\altaffilmark{3} 
}
\email{leahy@ucalgary.ca, morsink@ualberta.ca, yichou@astro.ncu.edu.tw }

\altaffiltext{1}{Department of Physics and Astronomy, University of Calgary,
Calgary AB, T2N~1N4, Canada}
\altaffiltext{2}{Department of Physics,
University of Alberta, Edmonton AB, T6G~2E1, Canada}
\altaffiltext{3}{Graduate Institute of Astronomy, National Central University, 
Jhongli 32001, Taiwan}

\begin{abstract}
The accreting millisecond pulsar XTE J1807-294 is studied through a pulse shape modeling analysis. 
The model includes blackbody and Comptonized emission from the one visible hot spot 
and makes use of the Oblate Schwarzschild approximation for ray-tracing.
We include a scattered light contribution, which accounts for flux scattered off an equatorial
accretion disk to the observer including time delays in the scattered light.  We give 
limits to mass and radius for XTE J1807-294 and compare to limits determined for
SAX J1808-3658 and XTE J1814-334 previously determined using similar methods. The
resulting allowed region for mass-radius curves is small but is consistent with
a mass-radius relation with nearly constant radius ($\sim$12 km) for masses between 
1 and 2.5 solar masses. 
\end{abstract}


\section{Introduction}
\label{s:intro}

One of the primary goals of neutron star astrophysics
is the observational determination of the mass-radius relation of neutron
stars, thus allowing the determination
of the equation of state (EOS) of cold supernuclear
density material \citep{Bhatta}. Modeling the pulsed light emitted from the surface of a
neutron star has the potential to attain this goal (e.g. Leahy, 2004a for mass radius
constraints for Her X-1). The
pulse shape encodes the emission region properties and the paths of photons 
traveling through the gravitational field of the spinning neutron star to the observer.
For a description of the Schwarzschild metric and of geodesics of light rays, see e.g.
Misner, Thorne and Wheeler (1973). 
For an introduction to emitting hot spots on neutron stars, see
e.g. Pechenick, Ftaclas and Cohen (1983).

Excellent candidates for a pulse-shape analysis are the accreting millisecond period
pulsars, where the energy released from accretion leads to
X-ray emission from the surfaces of the neutron stars.
SAX J1808.4-3658 (hereafter SAX J1808) was the first discovered
pulsar of this class \citep{WvdK98}. The
pulsations are thought to be produced from the accretion of plasma
funneled onto a hot spot at the neutron star's magnetic poles (see, e.g., Fig. 12
of \citet{GDB02}). 
Magnetohydrodynamic simulations for 
millisecond pulsars with weak magnetic fields
(e.g. see Kulkarni and Romanova 2005) show that the 
material flows from the inner edge of the accretion disk in some complicated
way down to the surface  to create a hot spot at some
position angle offset from the rotation axis. A further complication is
that pulse shapes of slowly rotating neutron star give evidence that their
magnetic fields are non-dipolar (e.g. Leahy and Li 1995, Leahy 2004b).
Thus we do not assume any fixed relation
between the spot latitude and inner accretion disk radius.
Spectral models \citep{Gil98,GDB02} provide strong evidence that the X-rays
correspond to blackbody emission from a spot on the star which
is then Compton scattered by electrons in a thin layer adjacent to 
the hot spot. 
Millisecond pulsars are low-luminosity systems (XTE J1807-294 in outburst 
was $10^{36} - 10^{37}$ erg/s, Falanga 2005).  SAX J1808-3658 rises from $<10^{32}$ erg/s 
in quiescence to 2-4 $\times 10^{36}$ erg/s in outburst (Hartman et al. 2008), so the accretion 
column is expected to be of small height above the surface. 
Other studies of pulse shapes of ms pulsars did not need to add emission above the surface to adequately model
the observed pulse shapes (e.g. Poutanen and Gierlinski 2003, Leahy, Morsink and Cadeau 2008).
We note also that emission from the sides of an accretion column produces a distinctive fan beam pattern in the 
observed pulse shape (e.g. Leahy, 2004b), which is not seen in the pulse shapes from
ms X-ray pulsars.
We make the assumption that the accretion column is of negligible height for simplicity.

A pulse-shape analysis requires tracing rays from the star to the 
observer. This is complicated by the rapid rotation of the star, 
since the metric and geodesics of a rotating neutron star must be
computed numerically. However, the Schwarzschild metric, describing
the exterior of a non-rotating star is simple and geodesics are 
simply found by quadrature. This leads to an often-used approximation,
 the Schwarzschild plus Doppler (S+D)
 (\citet{ML98};\citet{PG03}). In the S+D approximation, the 
gravitational field of the star is approximated as purely 
Schwarzschild, and all relativistic Doppler effects are added to the 
final expressions for flux as though the star's 
gravitational field were negligible. The validity of the 
S+D approximation was investigated \citep{Cadeau05,Cadeau07}
by computing the geodesics
of rapidly rotating neutron stars, using the exact numerical metric
to describe the gravitational field and the shape of the star.
\citet{Cadeau05,Cadeau07}
showed that it is necessary to include travel times of photons
from different parts of the star, but that it was sufficient to
use the Schwarzschild geometry to calculate geodesics as the
the effect of frame dragging is small. They also showed that
effect of the neutron star's oblate shape has a significant effect
on the light curves due to the ``shadowing'' effects caused by
the non-spherical shape of the star's surface.
The shape of the star needs to be included when modeling the
light curves of rapidly rotating neutron stars.  In 
\citet{MLCB07},
the oblate Schwarzschild (OS) approximation was introduced as a
simple, but sufficiently accurate calculation for modeling pulse shapes
of accreting millisecond pulsars.

The hot spot is assumed to be infinitessimal in size.
We have done many fits using an extended hot spot of various sizes and shapes (Morsink and 
Leahy, 2011). Within reasonable bounds (less than 3km spot diameter for a 10 km neutron star 
radius) the spot shape and size has negligible effect on the derived mass and radius of the
 neutron star. The same conclusion was reached by Lamb et al. (2009a) and Lamb et al (2009b)
who tested variation of spot shapes for spots with angular
sizes up to $45^\circ$.

The pulsations from two accreting ms pulsars have been analysed in order to constrain
the neutron star EOS. 
 \citet{PG03} modeled the pulse shape of SAX J1808
using the S+D approximation and gave constraints on the mass-to-radius ratio.
\citet{LMC08} applied the OS model, which includes oblateness and time-delays,
to give improved constraints on SAX J1808. Both studies only included data from
the 1998 outburst which skewed the analyses to very small stars. 
Subsequently this neutron star has gone through further
outbursts and the resulting pulse profiles were observed to vary over time
\citep{Har08}. This variation over time was included in the study
of \citet{ML11}, which separated data with different pulse shapes and modeled
these with common mass, radius and inclination angle parameters, but time-variable hot spot parameters.
The resulting mass-radius constraints (their Fig. 7) gave a moderate-mass compact star 
consistent with a number of mass-radius 
relations predicted by some common soft EOS. A similar multi-epoch analysis was carried out for
the millisecond pulsar XTE J1814-338 (hereafter XTE J1814) \citep{LMCC09}
which gave a high-mass large-radius star,
consistent with a stiff EOS. Studies of more accreting millisecond pulsars are needed to
further explore possible EOS, and to determine the overlap in the mass-radius plane for
different cases. Here we present a multi-epoch pulse shape study of XTE J1807-294 (hereafter XTE J1807). 

The accreting ms-Period pulsar XTE J1807 was discovered \citep{Markwardt2003} in February 2003 
when it went into outburst. The Rossi X-Ray Timing Explorer Proportional Counter Array (RXTE/PCA) 
observed the pulsar during most of its outburst, which lasted approximately 120 days. 
\citet{Zhang2006} noticed 4 events that they call ``puny'' flares, when the flux increases significantly, although
not as much as in a type I X-ray burst. \citet{Chou2008} identified 2 more ``puny'' flares and showed that 
the pulsar phase jumps discretely during the flare events (see Figure 3 of \citet{Chou2008}). Pulse
profiles can be constructed using data from periods between the flare events.

This paper is a summary of the analysis of the RXTE/PCA pulse shapes of XTE J1807. The data used in 
the models is described in Section~\ref{s:data}. 
Our model of the pulse 
profiles includes emission from a hot spot on the surface of the neutron star plus scattered emission from the
surface of the accretion disk. The model is described in detail in Section~\ref{s:models}.
  We use this model to put limits on the parameters of the pulse shape model,
including constraints on mass and radius of the pulsar in Section~\ref{s:results}.

\section{Data Analysis}
\label{s:data}

The observed pulse shapes of XTE J1807 are from RXTE/PCA observations of the 2003 outburst.
Since the pulse phase evolution is not continuous during the flare periods we consider data 
from times when the pulsar is not undergoing a flare.
We group the data into 3 non-flare epochs: Epoch 1 
corresponds to $12697 \le \hbox{TJD} < 12703$; Epoch 2 is $12716 < \hbox{TJD} < 12720$; 
and Epoch 3 is $12730 < \hbox{TJD} < 12759$. These 3 non-flare epochs are illustrated in Figure 2c of 
 \citet{Patruno2010} using circles; their figure shows a 4th non-flare epoch with very low flux 
that we do not use since it is too noisy.

All event times of non-flare epochs were first corrected to the barycenter of
the solar system and then
folded with the ephemeris, consisting of orbital Doppler effect and a
fourth-order polynomial pulse phase drift, proposed by  \citet{Chou2008}.
A time averaged pulse shape for each epoch was created by binning the event
phases into 32 bins.  All the pulse shapes were made separately for eight
narrow energy bands (bands 1 to 8, given by 1: 2-3.7 keV, 2:3.7-4.9 keV, 3:
4.9-5.7 keV, 4: 5.7-7.3 keV, 5: 7.3-9.0 keV, 6: 9.0-11.5 keV, 7: 11.5-14.8
keV and 8: 14.8-19.4 keV).

The $\chi^2$ test was applied to assess the difference in pulse shapes between different epochs
and different energy bands. 
To avoid differences in normalization we compared pulse shapes normalized to
unit amplitude.
Comparing the pulse shapes of a given energy band between epochs showed that, in all cases, the pulse shapes
were significantly different. As a result, we analyze all 3 epochs of data.

It is not computationally feasible to use all 8 energy bands in our modeling. It has been 
shown through spectral modeling \citep{PG03} of SAX J1808 that it is sufficient to use only
two energy bands when modeling an accreting ms pulsar's pulse profile. In particular, a
low-energy band is required to capture the low energy blackbody emission of the spot which 
typically peaks near 1 keV. The blackbody emission falls off quickly at higher energies, leaving
the higher energy bands almost completely dominated by the emission that has been Compton up-scattered
by the electron plasma lying overhead of the hot spot. For this reason we need to choose two
energy bands with pulse shapes that are as different as possible in our analysis.

In order to identify the energy bands to be used in our analysis, 
we compared pulse shapes for the same epoch between energy bands using the $\chi^2$ test.
For Epoch 1, the 2-3.7 keV pulse shape is significantly different 
from all other energy bands (reduced chi-squared: $\chi_r^2>1.9$). 
Most pairwise comparisons of pulse shapes showed significant differences. 
The exceptions for the 28 pairs for Epoch 1 were for comparisons between 
band $n$ and $n+1$ for $1 < n < 7$, which had $\chi_r^2<1.32$ (the value that corresponds to
less than a two sigma significance). This 
demonstrates the gradual change of pulse shape with energy and that band 1 is more different than the 
other bands. 

Epoch 2 had similar results: the same pattern of low $\chi_r^2$ was obtained for comparisons of neighboring 
energy bands, except with additions
of band 3 vs. band 5 and band 4 vs. band 6 included in the low $\chi_r^2$ set of pairs.
Epoch 3 also gave a similar pattern to Epoch 1, except that band 2 vs. band 3 ($\chi_r^2$=2.0) 
and band 3 vs. band 4 ($\chi_r^2$=1.35) pairs were not in the low $\chi_r^2$ set. 
The differences in the $\chi_r^2$ between the 3 epochs are in large part due to the differing amount of 
data averaged, and thus the size of the error bars (largest for
Epoch 2 and smallest for Epoch 3). 

In summary, the 2-3.7 keV band is the most different from the other 
energy bands, and the other energy bands only exhibit small changes in shape. This is consistent
with the spectral analysis of data from Epoch 3 by \citet{Falanga2005}, which finds a Comptonized
component in all energy bands, and an additional blackbody component in the low energy band.
Our pulse shape model was developed for this case of two spectral components in the low energy band and one
spectral component in a high energy band.
In order to include a significant amount of blackbody emission in our low energy band, we chose
the 2.0-3.7 keV band. For the high energy band we chose to combine bands 5 and 6, i.e. 7.3-11.5 keV.
This gave pulse shapes with smaller error bars and also avoids any possible contamination by iron
emission lines or the Blackbody component. Figs. 1, 2 and 3 show the observed pulse shapes
in these two energy bands for Epochs 1, 2 and 3, respectively.

\begin{figure}
\plotone{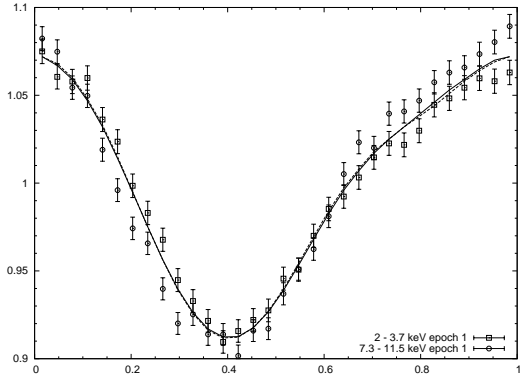}
\caption{Pulse shapes in two energy bands (squares 2.0 - 3.7 keV, circles 7.3 - 11.5 keV) for XTE J1807-294 
during epoch 1. The solid (low energy) and dashed (high energy) curves correspond to the best-fit model 
resulting from a simultaneous fit 
to the data in all three epochs. The best-fit model (see Table \ref{tab:1}) has parameter values 
$2M/R = 0.3$, $M =1.47 M_\odot$, $R = 14.7$ km and $\chi^2/$dof = 197/172.
}
\label{fig:1}
\end{figure}

\begin{figure}
\plotone{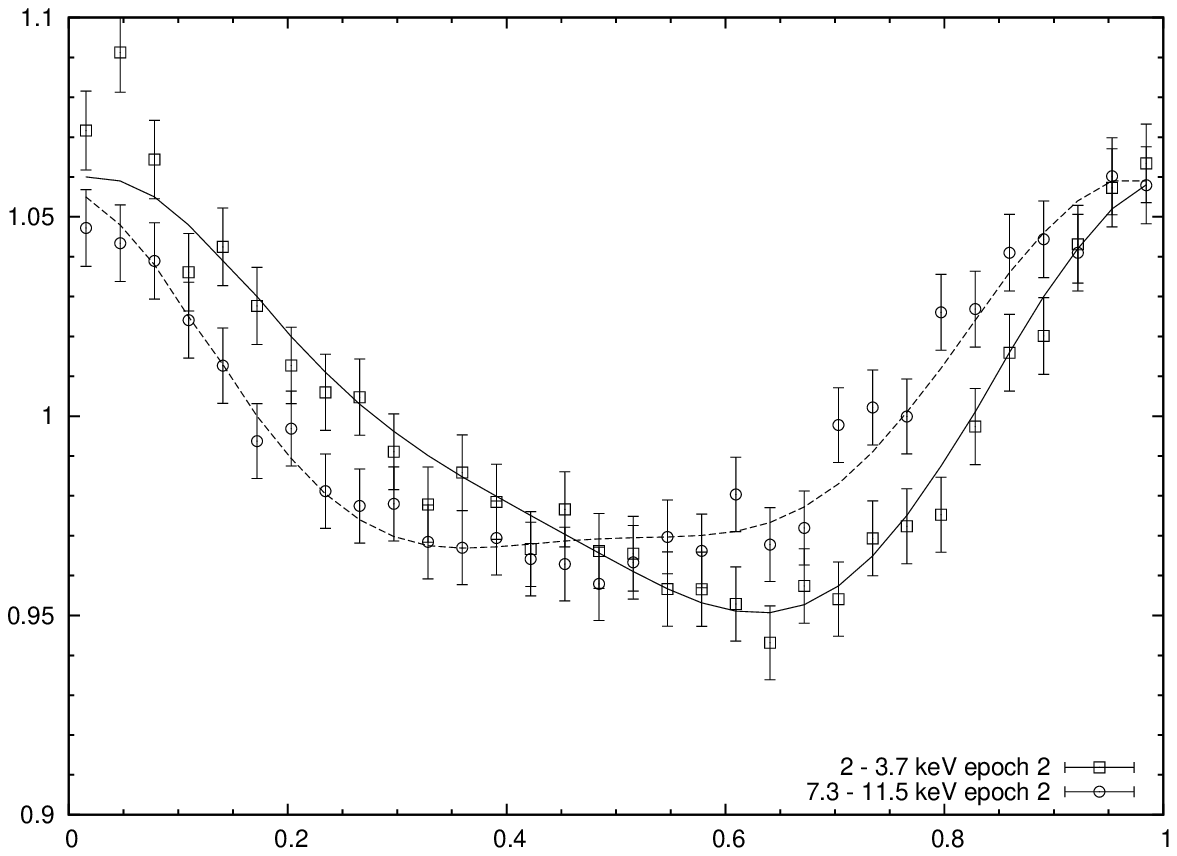}
\caption{Pulse shapes in two energy bands for XTE J1807-294 
during epoch 2. All symbols have the same meanings as in Figure 1.
}
\label{fig:2}
\end{figure}

\begin{figure}
\plotone{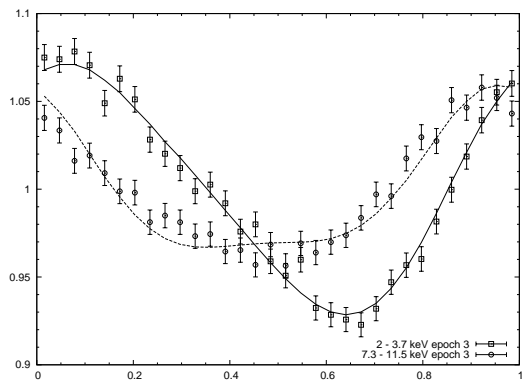}
\caption{Pulse shapes in two energy bands for XTE J1807-294 
during epoch 3. All symbols have the same meanings as in Figure 1.
}
\label{fig:3}
\end{figure}

\section{Pulse Shape Model}
\label{s:models}

The pulse shape models used in this analysis are similar to those used by  \citet{LMC08}  and \citet{ML11}
to model SAX J1808 and by \citet{LMCC09} to model XTE J1814. 
(These models are similar to that of \citet{PG03} for SAX J1808, but include time-delays
and oblateness, which are important, e.g. see \cite{MLCB07}). 
Our model computes pulse shapes in two energy bands: a high energy band dominated by Comptonized flux 
from the hot spot and a low energy band which includes both Comptonized and blackbody emission from
the hot spot.
As explained above, we make the assumption that the Comptonized emission from the accretion column is of negligible height above the surface for simplicity.

The hot spot model is described in detail in our previous work (e.g. \citet{ML11}, \citet{LMCC09},
and \citet{LMC08}). The previous work shows that the pulse shape depends very weakly on the hot spot 
size or shape (\citet{ML11}), so for the current work we use 
an infinitesimal spot. 
It is assumed that the Comptonized and Blackbody emissions both originate from the same
region on the surface of the star.
The emissivity of the Blackbody component is isotropic and the Comptonized component emissivity is
described by an analytic approximation, $I \propto 1 - a \cos\alpha$, where $\alpha$ is the angle between the
normal to the surface and the initial direction of the photon and $a$ is a free parameter. The low-energy
band includes both blackbody and Compton components, and the phase averaged ratio of the low-energy band
blackbody to Compton flux is represented by the parameter $b$.
The blackbody and Comptonized components have effective power-law indices $\Gamma$ 
over each observed energy band, which are determined by the observed X-ray energy spectrum. 
We made use of the spectral model of \citet{Falanga2005} to determine the values of $\Gamma$ in 
these two energy bands. 

To calculate a model pulse shape, the flux in each
phase-bin is computed using the oblate Schwarzschild approximation \citep{MLCB07}.
This accounts for the light-bending caused by gravity, the time-delays caused by
the finite travel time for photons across the star and the oblate shape of the 
rotating star. Since the light-bending depends only on the 
ratio of the neutron star's mass, M, to radius, R, at the spot location, the
star's gravitational radius, $2GM/Rc^2$, is kept fixed for a series of fits. 
(For simplicity, when referring to the gravitational radius, we use units with $G=c=1$.)
The lowest $\chi^2$ for a given $2M/R$ is found. 
There are eight free parameters for a fit to single pair of 2-energy band 
pulse shapes: 
R, $\theta$, $i$, $I_1$, $I_2$, $b$, $a$, and $\phi$.
The angle between the spot's centre and the spin axis is $\theta$, and the observer's
inclination angle to the rotation axis is $i$. $I_1$ and $I_2$ are the normalizations 
for the high energy and low energy pulse shapes and
the angle $\phi$ is an arbitrary rotation phase offset.
It is very unlikely that the spot and observer are in opposite hemispheres,
so we restrict both $\theta$ and $i$ to be less than $90^\circ$.  
The anisotropy is restricted to $0 < a < 1$. 

In this paper we restrict the models to only one visible spot. 
Prior to doing this we carried out fits to the data for the three epochs 
with a 2-spot model. The best fit 2-spot models in all cases had zero amplitude
for the second spot, indicating that no second spot is required by the observed
pulse shapes. This is consistent with the presence  of an accretion disk  which hides the antipodal spot.

In the current work, our goal is to simultaneously fit the two-energy-band pulse shapes 
from all three different epochs.  In 
the simultaneous fits, the parameters $M, R$, and $i$ are taken have the same
values. All other parameters are allowed to change their values.  This 
brings the number of free parameters to 20. Of these, six are normalizations and three are
phase offsets which are not of physical interest.

First we carried out test fits for a single epoch (Epoch 1), which had a small
number of free parameters (8). We found that the best fits had unacceptably large $\chi^2$
(205 for 56 degrees-of-freedom)
and had radius values that coincided with our imposed lower limit (4 km). This is similar 
behaviour to what we found for SAX J1808 \citep{ML11}, thus
we added a scattered light component (described below) and obtained much better
fits ($\chi^2$=72 for 54 degrees-of-freedom). Single epoch fits of Epoch 2 and Epoch 3 were
acceptable without need to add the scattered light component.
Joint fits of Epoch 1 with Epoch 2 or with Epoch 3 also had the same problem of high $\chi^2$
which was cured when a scattered light component was added for Epoch 1. 
A joint fit to Epoch 2 and Epoch 3 gave acceptable $\chi^2$ without the scattered light component.

The scattered light model is described in \citet{ML11}. It accounts for light 
that travels from the hot-spot to an optically thin plasma near the disk and then scatters into the line-of-sight
to the observer. Light-travel time delays and the Compton scattering angular
dependence are included. The formulae for the scattered light component are given in 
\citet{ML11}, specifically we use equation (9) with amplitude of $I_{sc}$
and distance of scattering region from the center of the star of $r_s$. 
This equation is the form for scattering from a disk in the equatorial plane of the neutron star.
Since the hot-spot is at high latitude on the neutron star,
the light from the hot spot that scatters to the observer is mainly from parts of the
disk that are far enough out that they are not moving relativistically. As shown in \citet{ML11},
 because the amplitude of the scattered light component is small (of order of a few \%), the model 
used for scattered light is adequate. 

With the scattered light component added to the Epoch 1 pulse shape model, the 
fits produce acceptable $\chi^2$ and reasonable parameter values. 
Since the simultaneous three
epoch fits produce tighter constraints on parameters than single epoch fits (or simultaneous two-epoch
fits), we present those results here.

\section{Results} 
\label{s:results}

The best-fit model that simultaneously fits the data from the three epochs is shown as solid and 
dashed curves in Figures \ref{fig:1}, \ref{fig:2}, and \ref{fig:3}. The best fit model corresponds to 
a neutron star with mass $M = 1.47 M_\odot$ and equatorial radius $R = 14.7$ km (gravitational radius $2M/R = 0.3$)
and $\chi^2/$dof = 197/172. Other parameter values for the best-fit model are shown in Table \ref{tab:1} in the
row for $2M/R = 0.3$. A series of fits keeping the gravitational radius (at the spot's location) fixed were performed,
and as can be seen from Table~\ref{tab:1}, the fits for other values of $2M/R<0.6$ are effectively as good as the fit
for $2M/R=0.3$ allowing for a wide range of allowed mass and radius values.

The parameter $\theta$ is the angle between the spin axis and the spot. In our initial fits of this data, 
we allowed the angle $\theta$ to vary between Epochs~1, 2 and 3. However, the best-fit solutions were always 
consistent with the value being the same during Epochs
2 and 3. For this reason, we simplified the analysis and introduced one single parameter $\theta_{23}$ to represent the 
angle between the spin axis and the spot during Epochs 2 and 3. The best-fit values for $\theta$  in all epochs are
small, consistent with the predictions made by \citet{Lamb2009}. The values of $\theta$ in Epoch~1 are smaller than 
that of Epochs~2 and 3 in Table~\ref{tab:1}.

\begin{figure}
\plotone{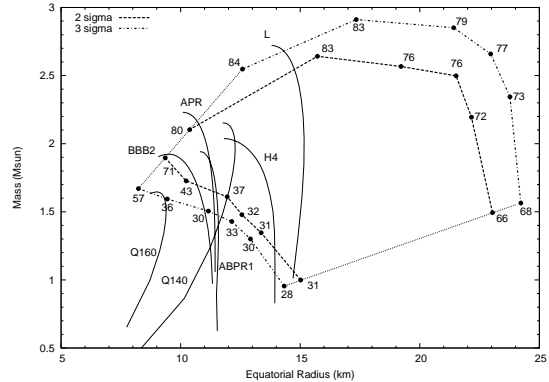}
\caption{Allowed Mass-Radius Region (at 2 and 3$\sigma$ confidence) for XTE J1807. The numbers
give the inclination angles for the models with M and R values at the black dots.
See the text for the descriptions of the equation of state curves (solid) used.
}
\label{fig:mr}
\end{figure}

Based on the lack of eclipses and the assumption that the compansion fills its Roche Lobe, \citet{Falanga2005} 
show that the binary's inclination angle, $i$, is  less than $83^\circ$. This restriction on the inclination angle
 was used in our fits. The resulting best-fit inclination angle is $38^\circ$ for $2M/R = 0.3$. However, as can 
be seen by the values shown in 
Table~\ref{tab:1}, a wide range of inclination angles is still allowed by the fits. The full range of allowed inclination 
angles is illustrated in Figure~\ref{fig:mr} , where the value of the inclination is shown as a number next to 
selected points on the $2\sigma$ and $3\sigma$ curves. The smallest inclination consistent with the data is $28^\circ$,
and all values up to limit of $83^\circ$ are allowed.

The parameter $a$ quantifies the anisotropy of the Comptonized emission, where $a=0$ corresponds to completely
isotropic emission and $a=1$ corresponds to ``fan'' beaming where emission is most intense for light emitted 
tangent to the surface. As in the case of the angle $\theta$, the anisotropy parameter was consistent with 
a constant value in Epochs 2 and 3, so we combined the anisotropy parameters for the last two epochs into
one parameter $a_{23}$. In Table~\ref{tab:1}, the best-fit values for the anisotropy in Epoch 1 ($a_1$) is small
$0.05 \le a_1 \le 0.08$ while the anisotropy in the latter epochs is large $0.61 \le a_{23} \le 0.76$. These values
can be understood by noting that in Epoch 1, there is very little difference between the low and high energy bands.
The low energy band includes both isotropic blackbody radiation as well as anisotropic Comptonized radiation,
while the high energy band only includes the Comptonized radiation. A difference in anisotropy will add relative 
phase shifts as well as pulse shape changes between the two energy bands. Hence, in order to keep the pulse shapes 
very similar (and to avoid a phase lag) the best-fit solution minimizes the anisotropy of the Compton scattered 
component, resulting in a small value for $a_1$. On the other hand, the high and low energy pulse shapes are 
significantly different in both of the latter epochs, allowing for large values of anisotropy. 

The parameter $b$ represents the ratio of blackbody radiation to Comptonized radiation in the soft energy band.
The pulsar was observed by the XMM \citep{Campana2003,Kirsch2004} and Integral  \citep{Falanga2005} satellites simultaneously 
for a period of
time (March 22, 2003) during Epoch 3. \citet{Falanga2005} fit the combined XMM, Integral and RXTE data (labeled rev-52 
in their paper) using a thermal Comptonization model along with blackbody emission from the disk. This spectral
model constrains the ratio of the spot's blackbody emission to Comptonized emission to be near 30\%, which is included 
in our models by adding a $\chi^2$ penalty for value of $b_3$ that lie outside of the \citet{Falanga2005} errorbars. 
In the first two epochs, the parameter $b$ is allowed to freely vary. 
In all of the fits, the ratio of Blackbody to Compton components in
the low energy band (5\% for Epoch 1, 10\% for Epoch 2 and 30\% for Epoch 3) 
increased as the source luminosity decreased.

The disk scattering model introduced by \citet{ML11} is a simple 2-parameter model for the scattering of light
emitted by the star by optically thin material near the disk. The two parameters are $I_{sc}$ the amplitude of 
the scattered flux (where the flux from the spot has an amplitude of 1) and $r_s$, the distance (in km) from the 
neutron star where the light is scattered. In these models, adding the scattered light component to the first
epoch improved the fits, however the addition of the scattered light to the later 2 epochs did not improve the fits.
For this reason, the scattered light was only included in Epoch 1. In the best fit models in Table~\ref{tab:1}, the
scattered light (in Epoch 1) corresponds to 2 - 7\% of the direct flux, and the scattering location is near 180 km.

Each best-fit solution provides an inclination angle and the neutron star's mass. Making use of the known
mass function \citep{Chou2008} the companion's mass can be calculated. Further, using the assumption that the companion is
filling its Roche lobe, its radius can be determined. For the best fit models shown in Table~\ref{tab:1},
the companion's mass is near $0.01 M_\odot$ and its radius is near $0.04 R_\odot$. According to the
models of finite temperature low-mass white dwarf stars computed by \citet{DB}, these values of mass
and radius for the companion is consistent with a hot C or O white dwarf, but not with a He white dwarf. 
The non-detection of the companion during the quiescent period, as reported by \citet{DAvanzo2009} is consistent with 
a very low mass white dwarf companion.

To explore limits on mass and radius for XTE J1807-294, we carried out an extensive grid of simultaneous
3-epoch fits in the mass-radius plane. The resulting 2 and 3 sigma limits on mass and radius are shown in 
Figure~\ref{fig:mr} with dashed and dot-dashed curves.  These curves are the result of a full variation of all the free 
parameters in the model. In our calculations we only performed fits for gravitational radii in the range 
$0.2 \le 2M/R \le 0.6$ that corresponds to the range of physical values expected for neutron stars. These two 
boundaries are shown as dotted lines in Figure~\ref{fig:mr}. At selected points (filled circles) along the 
$2\sigma$ and $3\sigma$ 
boundaries, the value of the binary's inclination angle (in degrees) is shown.

We find that high-mass or large-radius stars are favored. E.g. at the 3 sigma level, the (M,R) values 
(in units of $M_\odot$ and km) range from   
(1.67, 8.2) to (0.96, 14.3) along the lower-mass boundary and
(2.55, 12.6) to (2.91, 17.3) to (1.56, 24.2) along the higher-mass boundary. 
The larger radii prefered by these models arise due to the pronounced harmonics visible by 
eye in Figures~\ref{fig:1}-\ref{fig:3}. In this type of pulse-shape model, harmonics are
mainly due to the Doppler boosting effect, which requires a large velocity. A large 
velocity can come about from either a large radius, or a spot located near the equator. 
However, the small pulse-fraction tends to discriminate against larger values of $\theta$
as discussed by \citet{Lamb2009}.

For reference, mass-radius curves for stars spinning at a frequency of 191 Hz 
are shown for a few representative equations of state in Figure~\ref{fig:mr}.
All but one of the EOS chosen for this figure have the property that a mass of 
$1.93$ is allowed when the star spins at 317 Hz,
consistent with the mass measurement of PSR J1614-2230 \citep{Dem10}. The exception 
is  EOS Q160 (quark bag model, with a bag constant $B^{1/4} = 160 MeV$ 
\citep{Gle00}), which
is shown for illustrative purposes. The EOS used in Figure~\ref{fig:mr}
include  hadronic EOSs APR \citep{APR},  BBB2 \citep{BBB},
and L \citep{PPS76};
Hyperon EOS H4 \citep{LNO}; mixed phase quark-hadron EOS 
ABPR1 \citep{ABPR}; and a quark bag model EOS with a low value for the
bag constant, Q140. The data for XTE J1807-294 is only marginally consistent
with the Q160 EOS if the star is at the maximum allowed mass for this EOS.
The data is consistent with all other EOS shown in Figure~\ref{fig:mr}.
Clearly, any independent observation that constrains one of the 
free parameters (such as the inclination angle) would provide a 
stronger constraint on the EOS.

\section{Discussion} 
\label{s:discussion}

\begin{figure}
\plotone{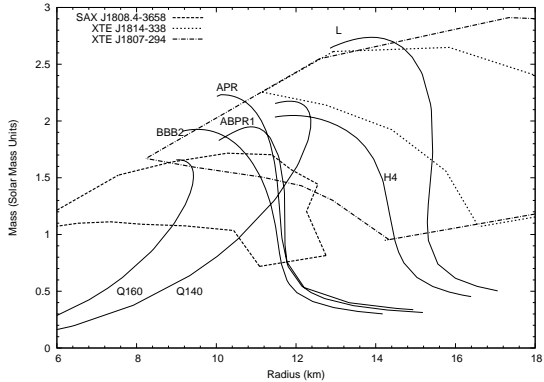}
\caption{Allowed $3 \sigma$ Mass-Radius regions for XTE J1807-294, SAX J1808 and XTE J1814-338. 
For reference mass-radius curves for stars spinning at 400 Hz are shown as solid lines. EOS
labels are the same as in Figure~\ref{fig:mr}.
}
\label{fig:all}
\end{figure}

A pulse-shape fitting analysis using the same procedure described in this paper has been performed for 
two other accreting ms-Period X-ray pulsars: SAX J1808 \citep{ML11} and 
XTE J1814 \cite{LMCC09}.  Figure~\ref{fig:all} shows the 3 $\sigma$ limits on
mass and radius for all three pulsars. The fits for SAX J1808 allow for 
only small masses and radii, while XTE J1814 allows for large masses and radii. The 
3 $\sigma$ region found in the present analysis for XTE J1807 encompasses almost all of
the region for XTE J1814, and overlaps with a small part of the region for SAX J1808.
The mass-radius curves shown in Figure~\ref{fig:all} correspond to stars 
spinning at a frequency of 400 Hz (the spin frequency of SAX J1808). Although the 
mass-radius curves change when the spin frequency changes, the differences 
are small enough that for the purpose of this figure, they do not alter the
results significantly.

As long as an EOS mass-radius curve has sections that pass through each of the 
star's 3 $\sigma$ allowed regions, it will be allowed by all 3 stars' data. 
None of the EOS curves in Figure~\ref{fig:all} possess this property. However,
an EOS that is somewhat stiffer than EOS APR and much softer than EOS L 
would pass through all 3 regions. For instance, an EOS mass-radius curve that
predicts a radius of 12 km for all masses smaller than about 2.4 $M_\odot$ 
would satisfy all 3 constraints.  Interestingly, \citet{steiner2010} found results consistent with this from an analysis of
3 type I bursters with photospheric radius expansion bursts, thermal emission from 3 transient LMXBs, 
and the isolated cooling neutron star RXJ1856-3754. They used a parameterized equation of state 
and a Monte-Carlo/Bayesian analysis to determine an EOS consistent with observations. They
found that the EOS 
is soft near nuclear saturation density, 
giving radii 11-12km for 1.4$M_\odot$, but the EOS stiffened 
at higher density
to give a maximum mass 1.9-2.3 $M_\odot$.
Their Figure 9 (right) summarized the probability distribution for M,R curves in the mass-radius plane,
and their allowed area coincides with the allowed region deduced above from the pulse shape analyses 
of SAX J1808, XTE J1814 and XTE J1807.

In summary, the
resulting allowed region for mass-radius curves is small but is consistent with
a mass-radius relation that has nearly constant radius ($\sim$12 km) for masses between  
1 and 2.5 solar masses. Observations of additional millisecond pulsars, that have pulse shapes sufficiently
bright for pulse shape modeling, are clearly needed to confirm and refine the results of the current work.

\acknowledgments
This research was supported by grants from the Natural Sciences and Engineering Research Council of
Canada. Y. Chou acknowledges partial support from Taiwan National Science Council
grants NSC NSC 100-2119-M-008-025.


\clearpage


\begin{deluxetable}{rrr|rrr|rr|rrr|rr|rr|r} \label{tab:1}
\tablecolumns{16}
\tabletypesize{\footnotesize} 
\tablewidth{0pc} 
\tablecaption{Best fit parameters for simulataneous fits to Epochs 1, 2 and 3. \label{tab:1}} 
\tablehead{ 
\multicolumn{3}{c}{Star}    &  \multicolumn{3}{c}{Geometry} &  \multicolumn{2}{c}{Anisotropy} &
\multicolumn{3}{c}{Spectrum} &
 \multicolumn{2}{c}{Scattering} & \multicolumn{2}{c}{Companion} & \colhead{}   \\ 
\colhead{$2M/R$}&\colhead{$M$}&\colhead{$R$}&\colhead{$\theta_1$}&\colhead{$\theta_{23}$}&
\colhead{$i$}&\colhead{$a_1$}&\colhead{$a_{23}$}&\colhead{$b_1$}&\colhead{$b_2$}&
\colhead{$b_3$}&\colhead{$I_{sc}$}&\colhead{$r_s$}&\colhead{$M_c$}&\colhead{$R_c$}&
\colhead{$\chi^2/$dof}\\
\colhead{}&\colhead{$M_\odot$}&\colhead{km}&\colhead{deg.}&\colhead{deg.}&
\colhead{deg.}&\colhead{}&\colhead{}&\colhead{}&\colhead{}&
\colhead{}&\colhead{}&\colhead{km}&\colhead{$M_\odot$}&\colhead{$R_\odot$}&
\colhead{}
}
\startdata
0.2 & 1.08 & 16.3 & 9   & 15 & 
41 & 0.06 & 0.64 & 0.027 & 0.121 & 
0.279 & 0.07 & 179 & 0.009 & 0.037  & 
198/172 \\ 
0.3 & 1.47 & 14.7 & 11 & 18 & 
38 & 0.06 & 0.61 & 0.018 & 0.122 & 
0.285 &  0.07 & 175 & 0.011 & 0.040  & 
197/172 \\ 
0.4 & 1.83 & 13.6 & 9   & 15 & 
50 & 0.06 & 0.65 & 0.017 & 0.124 & 
0.290 & 0.04 & 186 & 0.010 & 0.039  & 
198/172 \\ 
0.5 & 1.95 & 11.5 & 9   & 16 & 
56 & 0.08 & 0.66 & 0.015 & 0.126 & 
0.297 & 0.04 & 186 & 0.010 & 0.038  & 
198/172 \\ 
0.6 & 2.03 & 10.0 & 8   & 14 & 
79 & 0.05 & 0.76 & 0.013 & 0.127 & 
0.305 & 0.02 & 196 & 0.009 & 0.037  & 
201/172 
\enddata
\end{deluxetable}

\end{document}